\documentclass[manuscript]{aastex}
\usepackage{amsmath}
\usepackage{lscape}

\shorttitle{H$_2$CO and CCH images of IRAS 15398--3359} 
\shortauthors{Oya et al.}

\begin{document}
\title{A SUBSTELLAR-MASS PROTOSTAR AND ITS OUTFLOW OF IRAS 15398--3359 REVEALED BY SUBARCSECOND-RESOLUTION OBSERVATIONS OF H$_2$CO AND CCH} 
\author{Yoko Oya\altaffilmark{1}, Nami Sakai\altaffilmark{1}, Takeshi Sakai\altaffilmark{2}, 
		Yoshimasa Watanabe\altaffilmark{1}, Tomoya Hirota\altaffilmark{3}, \\ 
		Johan E. Lindberg\altaffilmark{4, 5}, Suzanne E. Bisschop\altaffilmark{4, 5}, 
		Jes K. J\o rgensen\altaffilmark{4, 5}, \\ 
		Ewine F. van Dishoeck\altaffilmark{6, 7}, and Satoshi Yamamoto\altaffilmark{1}}

\altaffiltext{1}{Department of Physics, The University of Tokyo, Bunkyo-ku, Tokyo 113-0033, Japan}
\altaffiltext{2}{Department of Communication Engineering and Informatics, 
			Graduate School of Informatics and Engineering, 
			The University of Electro-Communications, Chofugaoka, Chofu, Tokyo 182-8585, Japan}
\altaffiltext{3}{National Astronomical Observatory of Japan, Osawa, Mitaka, Tokyo 181-8588, Japan}
\altaffiltext{4}{Center for Star and Planet Formation, Natural History Museum of Denmark, University of Copenhagen, \O steer Voldgade 5-7, DK-1350 Copenhagen K., Denmark}
\altaffiltext{5}{Niels Bohr Institute, University of Copenhagen, Juliane Maries Vej 30, DK-2100 Copenhagen \O., Denmark}
\altaffiltext{6}{Leiden Observatory, Leiden University, P.O. Box 9513, 2300-RA Leiden, The Netherland}
\altaffiltext{7}{Max-Planck-Institut f\"{u}r Extraterrestrische Physik, Giessenbachstrasse 1, D-85748 Garching, Germany}

\received{2014 May 28}
\accepted{2014 September 16}

\begin{abstract}
Sub-arcsecond ($0\farcs5$) images of H$_2$CO and CCH 
line emission have been obtained in the $0.8$ mm band toward 
the low-mass protostar IRAS 15398--3359 in the Lupus 1 cloud 
as one of the Cycle 0 projects of the Atacama Large Millimeter/Submillimeter Array. 
We have detected 
a compact component concentrated in the vicinity of the protostar and 
a well-collimated outflow cavity extending along the northeast--southwest axis. 
The inclination angle of the outflow is found to be about $20\degr$, or almost edge-on, 
based on the kinematic structure of the outflow cavity. 
This is in contrast to previous suggestions of a more pole-on geometry. 
The centrally concentrated component is interpreted by use of a model of the infalling rotating envelope 
with the estimated inclination angle, 
and the mass of the protostar is estimated to be less than $0.09\ M_\odot$. 
Higher spatial resolution data are needed to infer the presence of a rotationally supported disk for this source, 
hinted at by a weak high-velocity H$_2$CO emission associated with the protostar. 
\end{abstract}

\section{Introduction}
Understanding the formation of disks around young low-mass protostars is an important target for studies of star formation. 
Although rotationally supported protostellar disks have been found in Class I sources, 
they have been reported in only a few Class 0 sources 
(Tobin et al 2012; Yen et al. 2013; Murillo et al. 2013; Lindberg et al. 2014). 
Even absence of such a disk down to $45$ AU is claimed for the Class 0 protostar NGC 1333 IRAS 2A (Maret et al. 2014; Brinch et al. 2009). 
Hence, disk formation in the Class 0 stage is still  controversial. 
This is because disks are generally difficult to identify in Class 0 sources 
due to the overwhelming emission from protostellar envelopes and outflows. 
Furthermore, the disk structure in Class 0 sources is expected to be small, 
and 
high spatial resolution and high sensitivity observations are essential. 

IRAS 15398--3359 is a low-mass Class 0 protostar in the Lupus 1 molecular cloud at a distance of $155$ pc (Lombardi et al. 2008). 
A molecular outflow was detected by single-dish observations 
of CO emission (Tachihara et al. 1996; van Kempen et al. 2009). 
Based on the relatively large overlap between the blue and red lobes of the outflow (CO $J=3$--$2$) 
observed with the James Clerk Maxwell Telescope ($\theta_{\rm HPBW} \sim 15^{\prime \prime}$), 
the outflow was thought to have a pole-on geometry (van Kempen et al. 2009). 
Mardones et al. (1997) observed H$_2$CO and CS lines toward this source 
to search for signs of infall without success. 
On the other hand, 
Kristensen et al. (2012) reported the presence of an inverse P-Cygni profile of the H$_2$O $1_{10}$--$1_{01}$ at $557$ GHz, 
indicating an infalling motion of the envelope on a scale of about $10^4$ AU. 
Recently, J\o rgensen et al. (2013) detected 
a ring structure of the H$^{13}$CO$^+\ J=4$--$3$  emission on $150$--$200$ AU scales. 
They proposed that the ring structure is caused by the destruction of HCO$^+$ 
through reactions with H$_2$O that is evaporated by the enhanced luminosity 
due to a recent accretion burst. 

IRAS 15398--3359 also shows peculiar chemical features in other ways. 
Various carbon-chain molecules such as CCH, C$_4$H, and CH$_3$CCH  
are detected 
in the vicinity of the protostar, 
which is characteristic of so-called 
warm carbon-chain chemistry (WCCC) sources (Sakai et al. 2008, 2009; Sakai \& Yamamoto 2013). 
Recently, 
the kinematic structure of another WCCC source, L1527, was resolved in Atacama Large Millimeter/Submillimeter Array (ALMA) observations, 
and was well reproduced by a model of an infalling rotating envelope (Sakai et al. 2014b). 
The radius of the centrifugal barrier, 
at which all the kinetic energy is converted to the rotational energy, 
was determined by comparing the observations with the model. 
From this radius, the protostellar mass was evaluated. 
Since both L1527 and IRAS 15398--3359 are WCCC sources, 
it is interesting to investigate whether the envelope of IRAS 15398--3359 has a similar structure. 
To explore when and how 
a rotationally supported disk 
is formed around protostars, 
a deep insight into the structure of the envelope of Class 0 protostars is useful. 
With these motivations, we conducted ALMA observations toward IRAS 15398--3359 in several molecular lines. 

\section{Observations}
Observations of IRAS 15398--3359 
were carried out with ALMA in Cycle 0 operations on 2012 December 31. 
Spectral lines of H$_2$CO and CCH were observed with the Band 7 receiver at frequencies of $349$--$352$ GHz, and $364$ GHz. 
The spectral line parameters are listed in Table $\ref{line}$. 
Twenty-five antennas were used in the observations, 
where the baseline length ranged from $13$ to $338$ m. 
The field center of the observations was 
($\alpha_{2000}$, $\delta_{2000}$) = ($15^{\rm h}43^{\rm m}02\fs3$, $-34\degr09^\prime07\farcs5$). 
The typical system temperature was $120$--$300$ K.  
The backend correlator was tuned to a resolution of $122$ kHz and a bandwidth of $469$ MHz, 
which corresponds to the velocity resolution of $0.1$ km s$^{-1}$ at $366$ GHz. 
J1517--243 was used for phase calibration every $12$ minutes. 
The bandpass calibration was carried out on J1256--057 for H$_2$CO and on J1924--292 for CCH, 
whereas the absolute flux density scale was derived from Mars and Titan, respectively. 
The data calibration was performed in the antenna-based manner 
and uncertainties are less than $10 \%$. 
Images were obtained by using the CLEAN algorithm. 
The continuum image was prepared by averaging line-free channels 
and the line maps were obtained after subtracting the continuum directly from the visibilities. 
The primary beam (half-power beam width) is $17^{\prime \prime}$. 
The total on-source time was $27$ minutes for H$_2$CO and $21$ minutes for CCH. 
The synthesized-beam size is 
$0\farcs57 \times 0\farcs42$ (P.A. $= 49\degr$) for the continuum image 
and $0\farcs60 \times 0\farcs44$ (P.A. $= 46\degr$) for the H$_2$CO image. 
The rms noise levels for the continuum and H$_2$CO emission are $0.001$ and $0.01\ {\rm Jy\ beam}^{-1}$, respectively. 
The continuum peak is: 
($\alpha_{2000}, \delta_{2000}) = (15^{\rm h} 43^{\rm m} 02\fs24, -34\degr 09^\prime 06\farcs7)$. 
The CCH data were combined with those taken in another observing program (2011.0.00628.S; PI: Jes J\o rgensen) 
with $15-16$ antennas. 
The signal-to-noise ratio (S/N) was improved from $13.7$ to $24.5$ with this addition. 
The synthesized beam of the combined CCH image is 
$0\farcs70 \times 0\farcs46$ (P.A. $= 72\degr$),  
and the rms noise levels is $0.015\ {\rm Jy\ beam}^{-1}$. 
We also detected the CH$_3$OH line 
as already published in a separate publication (J\o rgensen et al. 2013). 
We did not observe any other significant line features in the observed frequency range. 

\section{Results}
\subsection{Overall Distribution of H$_2$CO and CCH} 
\begin{landscape}
\begin{figure}
	\includegraphics[bb = 0 0 -500 450, scale = 0.58]{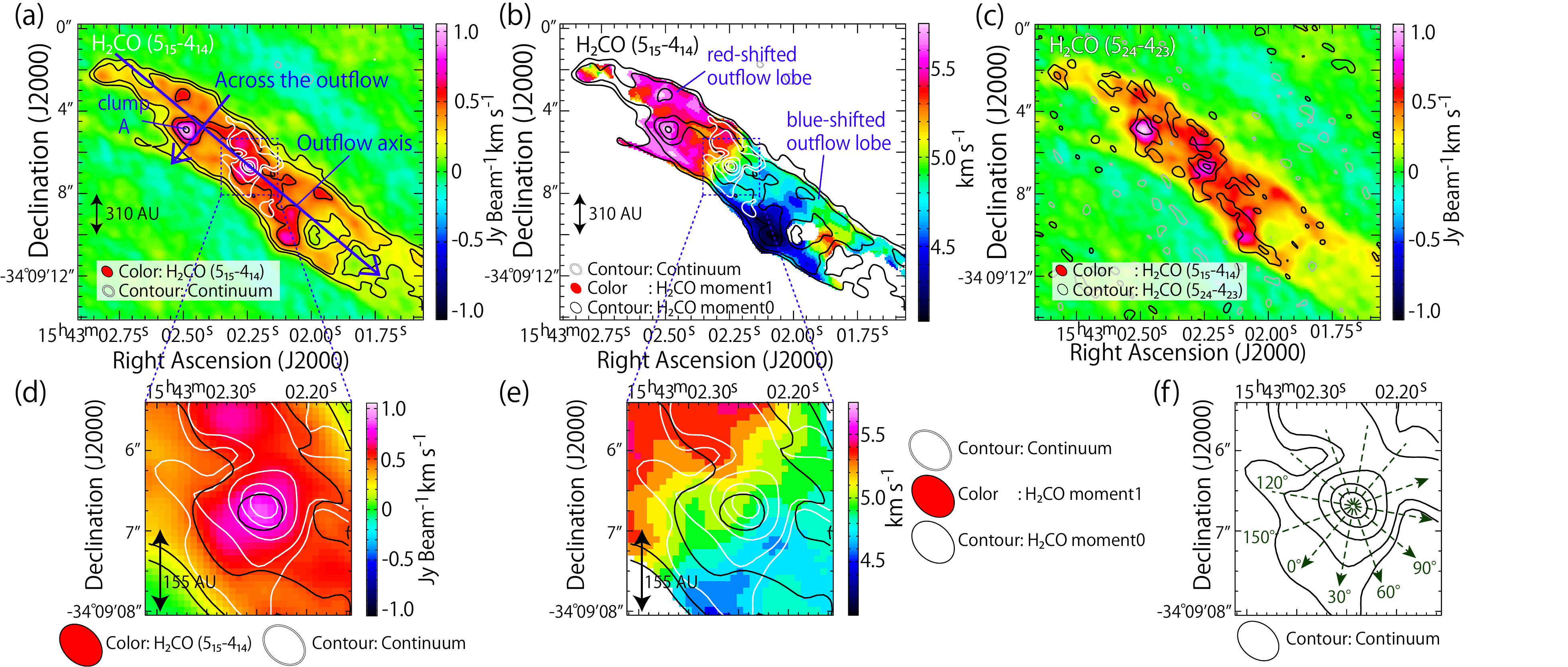}
	\caption{(a) Moment 0 map of H$_2$CO ($5_{15}$--$4_{14}$). 
				White contours represent the $0.8$ mm continuum distribution. 
				The blue arrows indicate the lines along which the PV diagrams 
				of Figures $\ref{PV_forI}$(a) and (b) are constructed. 
			(b) Moment 1 map of H$_2$CO ($5_{15}$--$4_{14}$). 
				The black and white contours are the same as in panel (a). 
			(c) Moment 0 map of H$_2$CO ($5_{24}$--$4_{23}$; contours) 
				superposed on the moment 0 map of H$_2$CO ($5_{15}$--$4_{14}$; color). 
				A few bright spots can be seen in both maps. 
			(d) A zoom of the central part of panel (a). 
				The black and white contours are the same as in panel (a). 
			(e) A zoom of the central part of panel (b). 
				The black and white contours are the same as in panel (a). 
			(f) A zoom of the central part of the continuum map. 
				The black broken arrows indicate the lines along which the PV diagrams of 
				Figures $\ref{PV2x2}$, $\ref{H2CO_PVdiff}$, and $\ref{H2CO_PV6}$ are prepared. 
			\label{H2CO_moment}}
\end{figure}
\end{landscape}

Figure $\ref{H2CO_moment}$(a) shows the moment 0 (integrated intensity) map of 
H$_2$CO ($5_{15}$--$4_{14}$). 
The most prominent feature is a well-collimated outflow extending symmetrically from the protostar 
along a northeast--southwest axis. 
The outflow has a very straight-wall structure, 
as in the case of the HH46 outflow (Arce et al. 2013). 
The northeastern lobe of the outflow is redshifted, 
while the southwestern lobe is blueshifted, as shown in the moment 1 (velocity field) map of Figure $\ref{H2CO_moment}$(b). 
In these observations, 
the emission extending over scales of $12^{\prime \prime}$ or larger is not reliable due to the lack of short baselines. 
Although the outflow size looks compact (about $8^{\prime \prime}$ for each lobe), 
the emission of the outer part may be resolved out or may be weak due to insufficient excitation conditions. 
The apparent width of the outflow is $4^{\prime \prime}$ at a distance of $8^{\prime \prime}$ from the protostar. 
Figure $\ref{H2CO_moment}$(c) shows the moment 0 map of a high-excitation line of 
H$_2$CO ($5_{24}$--$4_{23}$). 
The distribution of H$_2$CO ($5_{24}$--$4_{23}$) is essentially similar to that of H$_2$CO ($5_{15}$--$4_{14}$), 
although the S/N ratio of the former is rather poor. 
In these maps, a bright knot (Clump A in Figure $\ref{H2CO_moment}$(a)) 
can be seen in the redshifted lobe, 
which could be a shocked region caused by an impact of the outflow with dense clumps in a surrounding cloud. 

\begin{figure}
	\includegraphics[bb = -150 0 100 0, scale = 0.5]{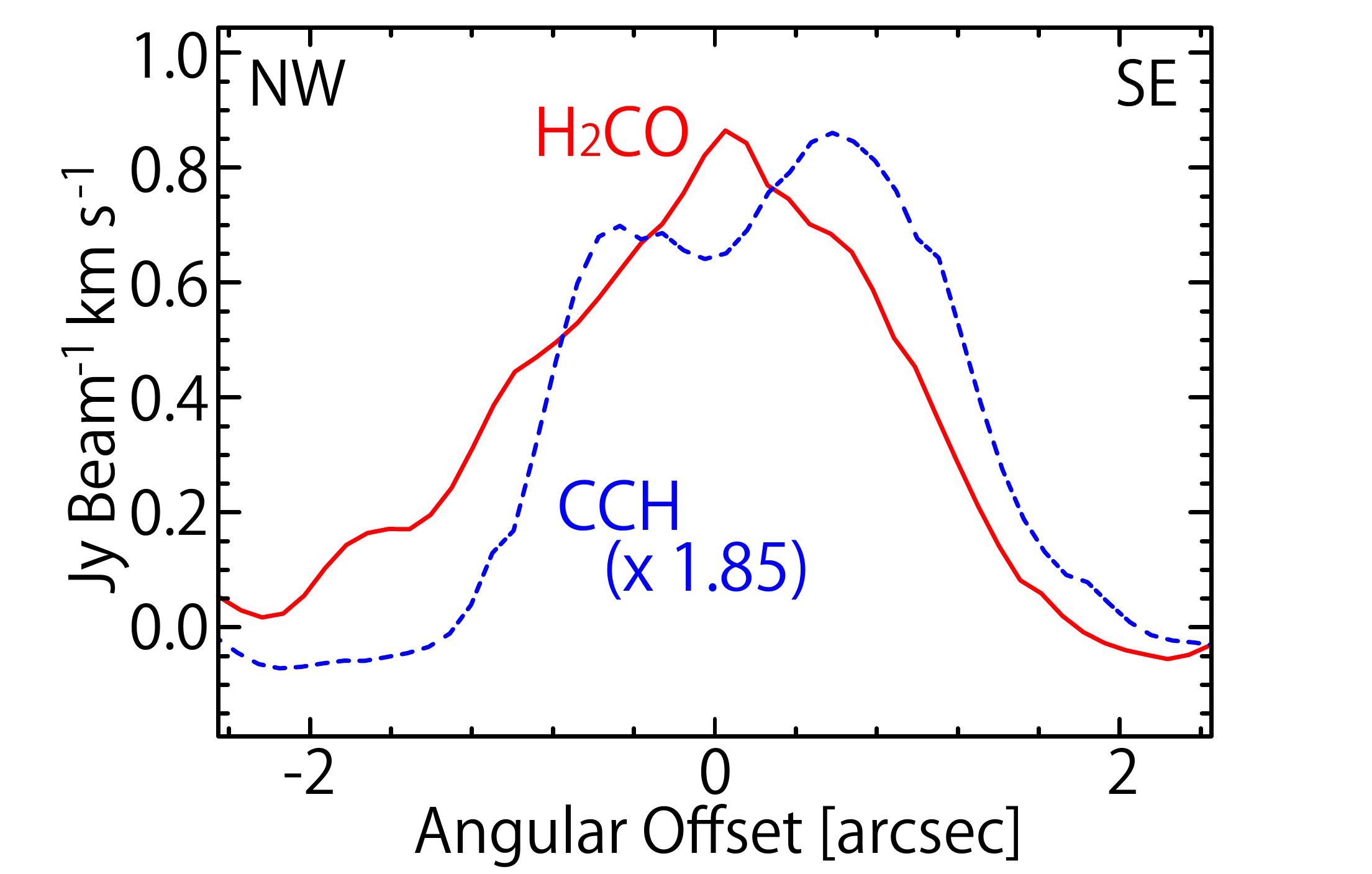}
	\caption{Intensity profiles of H$_2$CO ($5_{15}$--$4_{14}$) and CCH ($F_2$) 
			along the line perpendicular to the outflow axis 
			(``$0\degr$" shown in Figure $\ref{H2CO_moment}$(f)). 
			\label{I-Profile}}
\end{figure}

\begin{figure}
	\includegraphics[bb = 0 0 100 800, scale = 0.7]{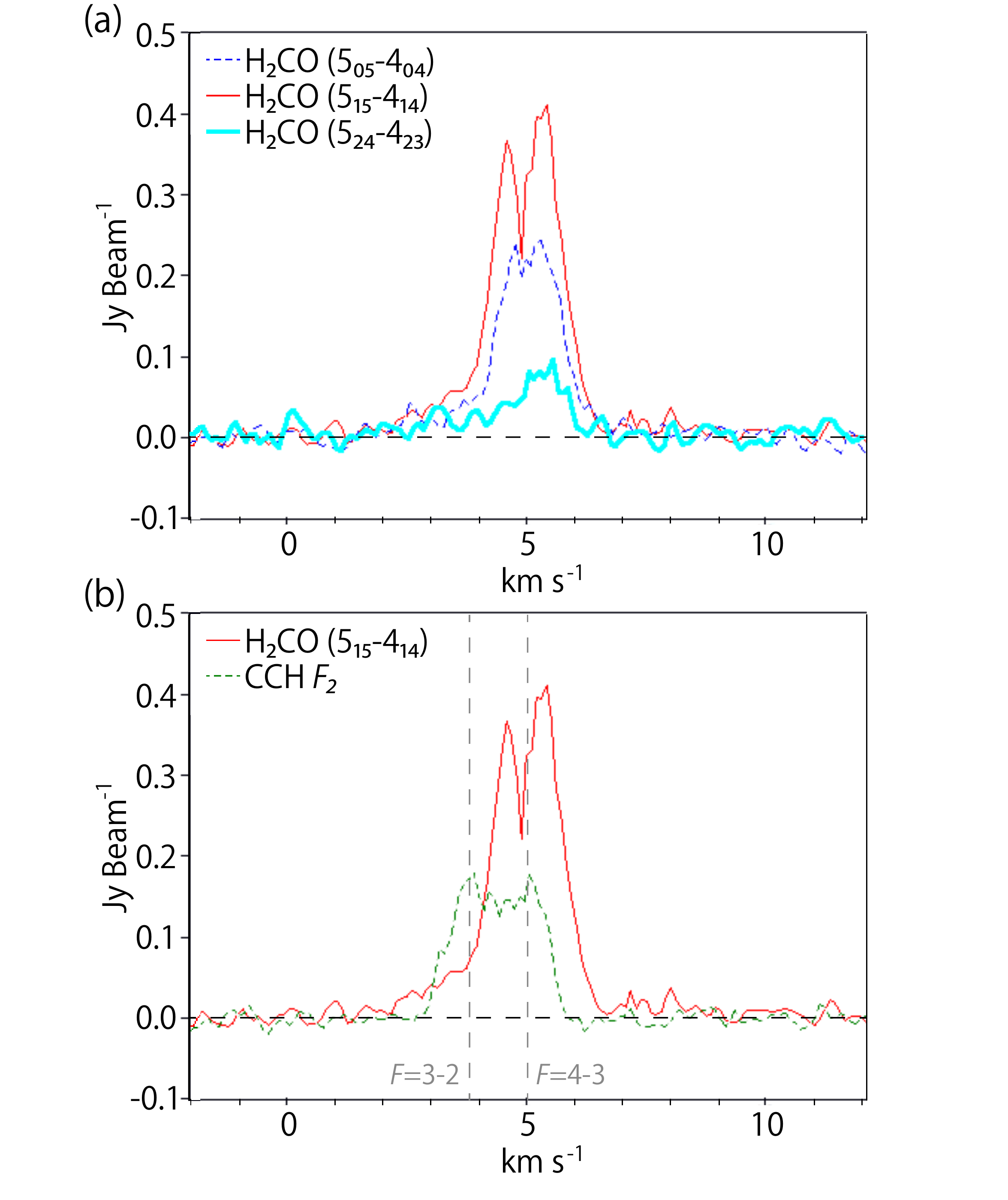}
	\caption{Spectra of H$_2$CO ($5_{05}$--$4_{04}$, $5_{15}$--$4_{14}$ and $5_{24}$--$4_{23}$) and CCH ($N$=4--3, $J$=7/2--5/2, $F$=4--3 and 3--2) toward the protostar position. 
			Two vertical dotted lines in (b) represent the systemic velocities for the two hyperfine components of CCH.}
	\label{spectra}
\end{figure}

In addition to the outflow feature, 
a centrally concentrated component with a single-peaked distribution can also be recognized, 
as shown in a zoom of the central part of Figure $\ref{H2CO_moment}$(a) (Figure $\ref{H2CO_moment}$(d)). 
A blowed-up of the moment 1 map is also shown in Figure $\ref{H2CO_moment}$(e). 
The approximate extent of the central component is estimated to be about $2^{\prime \prime}$ in diameter 
based on the intensity distribution along the line perpendicular to the outflow (Figure $\ref{I-Profile}$), 
which corresponds to $310$ AU. 
Figure $\ref{spectra}$ shows the spectral line profiles of H$_2$CO and CCH toward the protostar position averaged over the synthesized beam. 
The line width of H$_2$CO is as narrow as $2\ {\rm km\ s}^{-1}$ even toward the protostar position.
Using the RADEX program (van der Tak et al. 2007) 
to fit the intensity of the two temperature sensitive lines of para-H$_2$CO ($5_{05}$--$4_{04}$, $5_{24}$--$4_{23}$), 
the column density of H$_2$CO and the kinetic temperature toward the protostar 
are estimated to be $3 \times 10^{13}\ {\rm cm}^{-2}$ and $36$--$38\ {\rm K}$ 
on the assumption that 
H$_2$ density is $1 \times 10^7$--$1 \times 10^8\ {\rm cm}^{-3}$ 
and the line width is $1.8\ {\rm km\ s}^{-1}$. 
In addition, 
the ortho/para ratio is estimated to be $2.8$ 
with the intensity of an ortho-H$_2$CO line ($5_{15}$--$4_{14}$). 
The optical depths for these lines are 
$0.27$ ($5_{05}$--$4_{04}$), $0.06$ ($5_{24}$--$4_{23}$), $0.53$ ($5_{15}$--$4_{14}$), 
respectively, therefore the lines are not opaque. 

\begin{figure}
	\includegraphics[bb = 0 0 100 400, scale = 0.73]{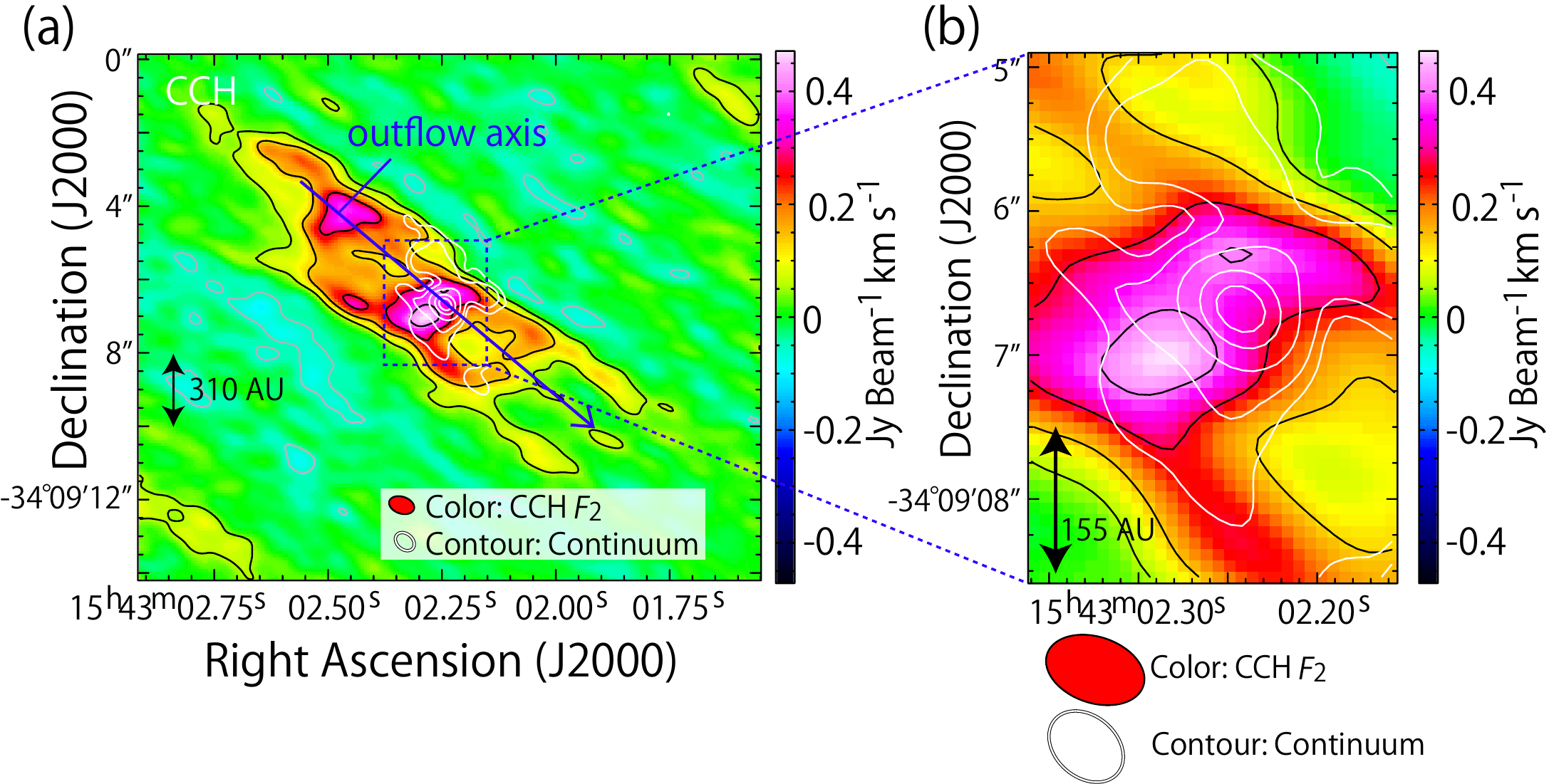}
	\caption{(a) Moment 0 map of CCH ($N$=4--3, $J$=7/2--5/2, $F$=4--3 and 3--2; $F_2$). 
				White contours represent the $0.8$ mm continuum distribution, 
				which is the same as in Figure $\ref{H2CO_moment}$(a). 
				The blue arrow indicates the line along which the PV diagram of Figure $\ref{PVoutflow}$ is prepared. 
			(b) A zoom of the central part of panel (a). 
				The black and white contours are the same as in panel (a). 
			\label{CCH_moment}}
\end{figure}

Figure $\ref{CCH_moment}$(a) shows the moment 0 map of CCH ($N$=4--3, $J$=7/2--5/2, $F$=4--3 and 3--2), 
while Figure $\ref{CCH_moment}$(b) is a zoom of its central part. 
The outflow cavity is prominent 
and the centrally concentrated component can also be seen. 
The CCH distribution is more extended along the southeast--northwest axis 
than the H$_2$CO distribution. 
It seems to have a slight dip toward the protostar position 
as shown in the intensity profile along the line perpendicular to the outflow axis (Figure $\ref{I-Profile}$). 
This feature of CCH is consistent with that reported by J\o rgensen et al. (2013). 
Although the spectrum of CCH at the protostar position is complicated because of the two hyperfine components (Figure $\ref{spectra}$), 
the line width of each hyperfine component of CCH is $2\ {\rm km\ s}^{-1}$ or less, 
as in the case of H$_2$CO. 

\subsection{Outflow} 
\begin{figure}
	\includegraphics[bb = 0 0 100 250, scale = 0.65]{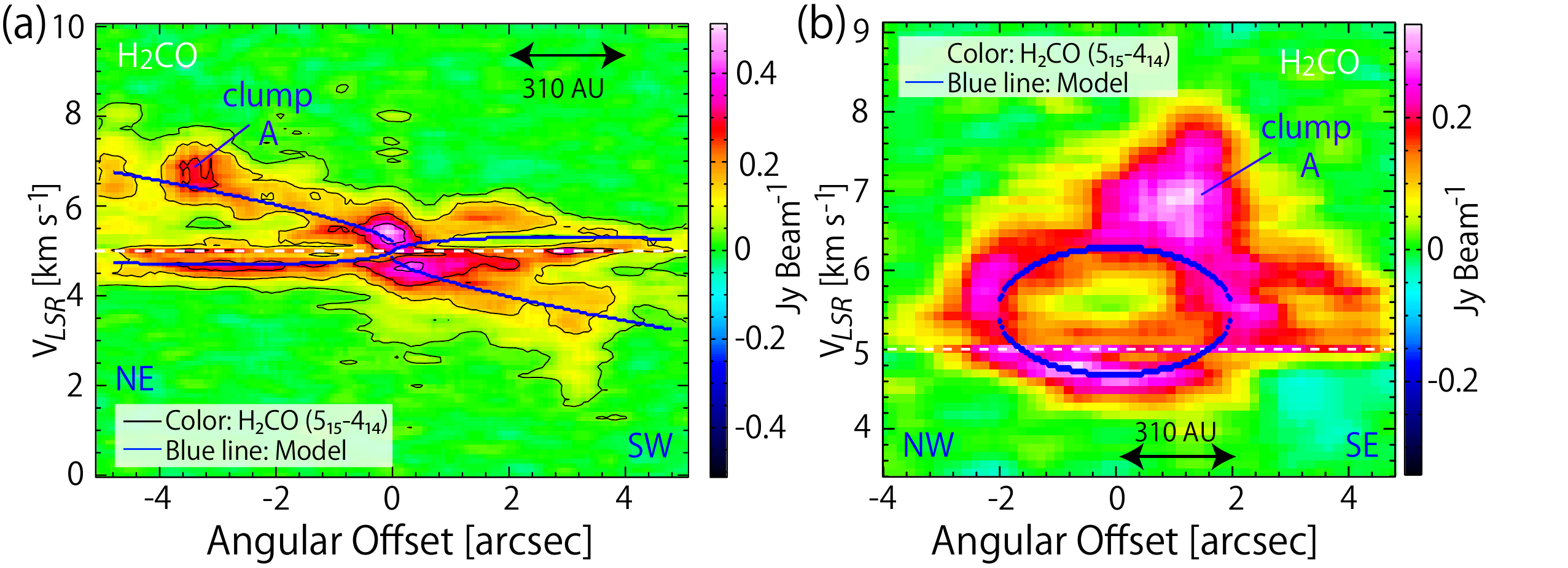}
	\caption{PV diagrams of H$_2$CO ($5_{15}$--$4_{14}$) 
				(a) along the outflow axis 
				and (b) across the outflow axis shown in Figure $\ref{H2CO_moment}$(a). 
				The blue lines show the best model with an inclination angle of $20\degr$. 
				The white broken lines represent the systemic velocity. 
				\label{PV_forI}}
\end{figure}

We first analyze the outflow feature observed in the H$_2$CO ($5_{15}$--$4_{14}$) line. 
Figure $\ref{PV_forI}$(a) shows the position--velocity (PV) diagram of the H$_2$CO ($5_{15}$--$4_{14}$) line emission 
along the outflow axis through the protostar position shown in Figure $\ref{H2CO_moment}$(a). 
The outflow has a redshifted component extending to the northeast from the protostar 
and a blueshifted component extending to the southwest from the protostar. 
Since the redshifted and blueshifted lobes of the outflow show little overlap 
with each other around the protostar,  
the outflow axis seems to be close to the plane of the sky, 
indicating that the disk/envelope geometry is almost edge-on. 
The highest velocity at a certain distance to the protostar linearly increases as a function of the distance, 
as often observed for outflow cavities (e.g., Lee et al. 2000; Arce et al. 2013). 
The intense knot in the redshifted component corresponds to Clump A in Figure $\ref{H2CO_moment}$(a). 
In addition to the high-velocity component, 
another velocity component can be seen around the systemic velocity ($\sim 5$ km s$^{-1}$; Sakai et al. 2009). 
This component is slightly blueshifted on the northeastern side of the protostar 
and redshifted on the southwestern side of the protostar, 
which is the reverse case compared to the high-velocity component. 

Figure $\ref{PV_forI}$(b) is 
the PV diagram of H$_2$CO ($5_{15}$--$4_{14}$) 
along a line perpendicular to the outflow axis in the redshifted lobe indicated in Figure $\ref{H2CO_moment}$(a). 
The PV diagram shows an elliptic feature with 
a knot-like distribution at the higher-velocity range. 
This knot-like structure corresponds to Clump A in Figure $\ref{H2CO_moment}$(a). 
The gas in the cavity wall seems to be expanding. 

We employ the standard model of an outflow cavity from Lee et al. (2000) 
to analyze the observed geometrical and kinematical structures of the outflow, 
where the outflow cavity is assumed to have a parabolic shape 
and its velocity is proportional to the distance to the protostar: 
\begin{equation}
	z = C R^2, \quad v_R = v_0 \frac{R}{R_0}, \quad v_z = v_0 \frac{z}{z_0} 
\end{equation}
where $z$ denotes the distance to the protostar along the outflow axis, 
and $R$ the radial size of the cavity perpendicular to $z$. 
$R_0$ and $z_0$ are both normalization constants, and are set to be $1^{\prime \prime}$.  
$C$ and $v_0$ are free parameters. 
The best results are obtained with an inclination angle of $20\degr$, as shown by the blue lines in Figures $\ref{PV_forI}$(a) and (b). 
When the inclination angle is higher than $30\degr$ or less than $10\degr$, 
the model does not reproduce the observations well 
with any values of $C$ and $v_0$. 
Hence, the inclination angle is determined to be $20\degr \pm 10\degr$, 
where the quoted error is the estimated limit of error based on the above analysis. 
The derived parameters are: $C = 0.8$ arcsec$^{-1}$ and $v_0 = 0.38$ km s$^{-1}$ 
for an inclination angle of $20\degr$. 
As mentioned in Section 1, van Kempen et al. (2009) 
reported an inclination angle of $75\degr$. 
This discrepancy seems to originate from 
the limited spatial resolution of their data. 
Another possibility is that the outflow direction on small scales is different from that on larger scales 
(e.g., Sakai et al. 2012; Kristensen et al. 2013; Y\i ld\i z et al. 2012; Mizuno et al. 1990). 
On the other hand, recent Atacama Pathfinder EXperiment observations of CO $J$=6--5 indicate an inclination angle of $20\degr$ (U.A., Y\i ld\i z et al. submitted to A\&A), 
which is consistent with our result. 

\begin{figure}
	\includegraphics[bb = -30 0 100 200, scale = 1.0]{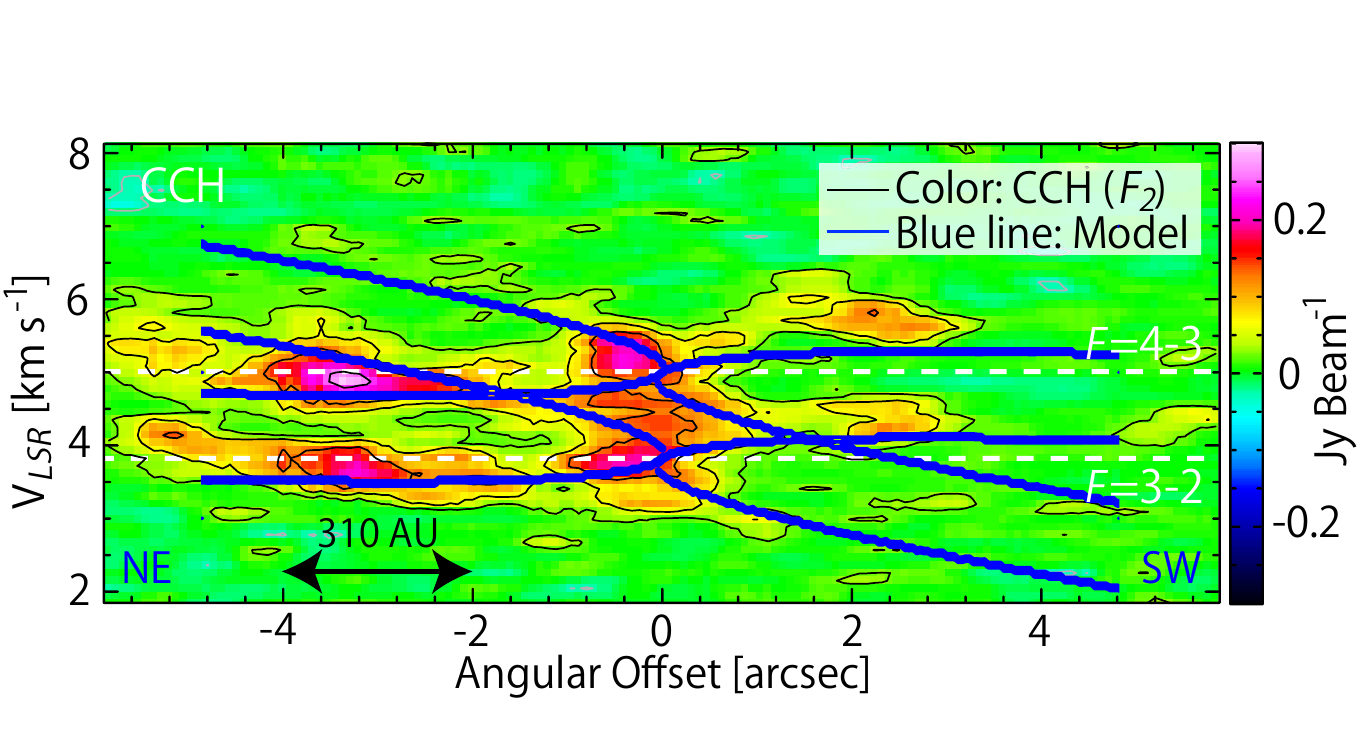}
	\caption{PV diagram of CCH ($F_2$) along the outflow axis (see Figure $\ref{CCH_moment}$(a)). 
			The blue lines show the model with an inclination angle of $20\degr$ 
			which is the best model for the H$_2$CO case. 
			The white broken lines represent the systemic velocities. 
			\label{PVoutflow}}
\end{figure}

Figure $\ref{PVoutflow}$ shows the PV diagram of CCH ($F_2$) along the outflow axis through the protostar position. 
The PV diagram is complicated, because the two hyperfine components are blended 
with a separation of only $1.2\ {\rm km\ s}^{-1}$. 
In comparison with the H$_2$CO case, 
the low-velocity component close to the systemic velocity 
is relatively bright in comparison with the high-velocity component. 
This means that the CCH emission from the outflow part primarily traces 
the compressed ambient gas around the outflow cavity rather than the entrained outflow gas. 
CCH may be formed by gas-phase reactions in dense photodissociation region layers 
(e.g. Jansen et al. 1995a, 1995b; Sternberg \& Dalgarno 1995).  

\subsection{Protostellar Envelope}
As shown in Figures $\ref{H2CO_moment}$(d) and $\ref{CCH_moment}$(b), 
there is a centrally concentrated component in the H$_2$CO and CCH emission. 
The distributions of C$^{34}$S, C$^{17}$O, and CH$_3$OH also have 
such a component (J\o rgensen et al. 2013). 
Here, we investigate the kinematics of this component. 
A zoom of the moment 1 map of H$_2$CO (Figure $\ref{H2CO_moment}$(e)) is dominated by the overwhelming outflow motion. 
However, a slightly skewed feature around the continuum peak is marginally recognized. 
In order to reveal the motion near the protostar more carefully, 
we prepared PV diagrams. 

\begin{figure}
	\includegraphics[bb = 0 0 100 600, scale = 0.5]{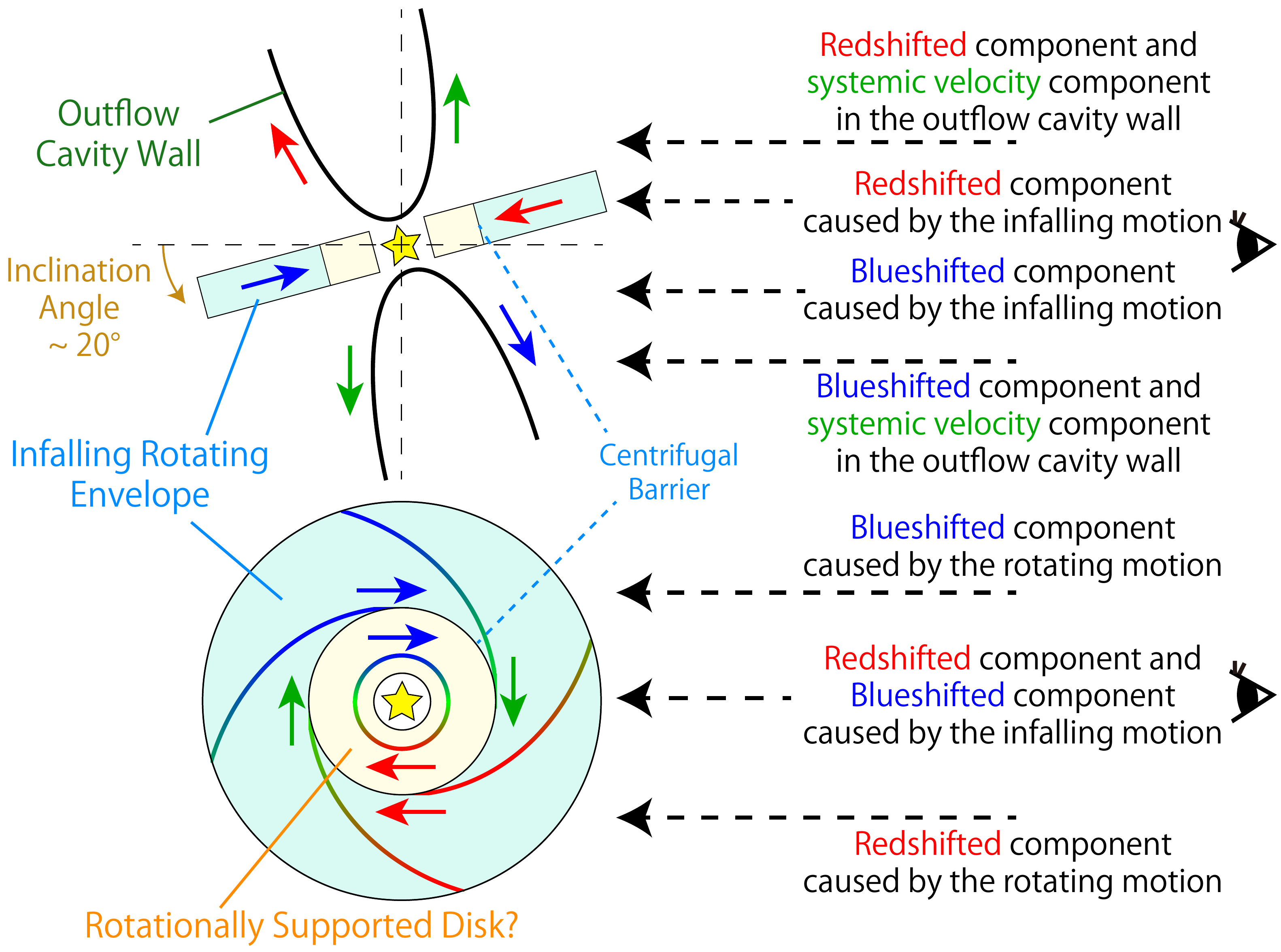}
	\caption{Model of the infalling rotating envelope and the rotationally supported disk. 
			An observer sits on the right hand side. 
			The black broken arrows represent the lines of sight. 
			The disk/envelope geometry is almost edge-on with an inclination angle of $20\degr$. 
			\label{model}}
\end{figure}

Figures $\ref{PV2x2}$(a) and (b) show the PV diagrams of H$_2$CO ($5_{15}$--$4_{14}$) along the two lines 
centered at the protostar position shown in Figure $\ref{H2CO_moment}$(f). 
Figure $\ref{PV2x2}$(a) is along the axis perpendicular to the outflow axis, 
which is shown by a broken arrow labeled as ``$0\degr$" in Figure $\ref{H2CO_moment}$(f). 
Figure $\ref{PV2x2}$(b) is along the outflow axis, 
which is shown by a broken arrow labeled as ``$90\degr$" in Figure $\ref{H2CO_moment}$(f). 
In Figure $\ref{PV2x2}$(a), we can see 
a marginal trend that the intensity peaks in the redshifted and the blueshifted velocity ranges are 
in the southeastern and northwestern sides of the protostar position, respectively, 
although a rotation signature is not obvious in the centrally concentrated component. 
%
On the other hand, 
two emission peaks are seen in Figure $\ref{PV2x2}$(b), 
one of which is redshifted in the northeastern side of the protostar position 
and the other is blueshifted in the southwestern side. 
This systematic velocity gradient in Figure $\ref{PV2x2}$(b) is the same case of that in the outflow, 
but the modest velocity gradient 
in the vicinity of the protostar 
is difficult to attribute to the outflow, 
according to our outflow model. 
Considering the outflow direction and the inclination angle, 
the outflow-envelope structure of this source is expected to be like a schematic illustration 
shown in the upper panel of Figure $\ref{model}$. 
Therefore, 
the velocity gradient could be a signature of an infalling motion in the envelope rather than outflow. 

An infalling/rotating signature is not clearly seen in the PV diagrams for CCH ($F_2$) 
(Figures $\ref{PV2x2}$(c) and (d)), 
in contrast with the well-studied low-mass Class 0 protostar L1527 (Sakai et al. 2014a, 2014b) at a distance of $140$ pc (Torres et al. 2007).  
In 
L1527, 
whose protostellar mass and inclination angle are $0.18\ M_\odot$ and $5\degr$, respectively, 
the velocity shift from the systemic velocity
observed for the CCH line is $1.8$ km s$^{-1}$ at a radius of $100$ AU from the protostar (Sakai et al. 2014a). 
On the other hand, 
the velocity shifts from the systemic velocity are as small as 
$1$ km s$^{-1}$ and $0.7$ km s$^{-1}$ for H$_2$CO and CCH, 
respectively, in IRAS 15398--3359 (Figure $\ref{PV2x2}$), 
despite the low inclination angle of this source ($20\degr$). 
Although IRAS 15398--3359 is similar to L1527 in its large-scale ($\sim$ a few $1000$ AU) chemical composition (Sakai et al. 2009), 
the infalling/rotating motion is not very clear. 

In this observation, 
we resolve structure down to $78$ AU ($0\farcs5$) around the protostar. 
Nevertheless the observed line width is narrow. 
This means that 
the Doppler shift due to infalling/rotating motions around the protostar should be small. 
In principle, the small velocity shift could be explained 
if the dense gas were not associated with the protostar, 
that is, 
if it were mostly present in the outflow. 
However, this possibility seems unlikely 
because 
the outflow motion cannot well explain the velocity gradient in Figure $\ref{PV2x2}$(b) for H$_2$CO, 
as mentioned above. 
Hence, the small velocity shift likely implies a low protostellar mass. 
The upper limit of the central mass is roughly estimated from the maximum velocity shifts from the systemic velocity. 
Under the energy conservation law, 
the central mass $M$ can be represented in terms of 
the infalling velocity ($v_{\rm infall}$) and the rotation velocity ($v_{\rm rotation}$) 
at the distance $r$ to the protostar as: 
\begin{equation}
	M = \frac{r}{2G} \left( v_{\rm infall}^2 + v_{\rm rotation}^2\right). 
\end{equation}
By use of this relation, 
the central mass can roughly be estimated to be smaller than $0.09\ M_\odot$ 
with a maximum velocity of less than $1$ km s$^{-1}$ at $0\farcs5$ ($78$ AU) from the protostar. 
Here, we assume as a robust case 
that the infalling velocity and the rotating velocity are both 
the observed maximum value of $1\ {\rm km\ s}^{-1}$ in Figure $\ref{PV2x2}$(a). 
In the case of no rotation (free fall), 
the upper limit to 
the central mass is estimated to be about $0.04\ M_\odot$. 

\subsection{Comparison with an Envelope Model}
In the case of L1527, which has similar chemical features to IRAS 15398--3359, 
the toy model of the infalling rotating envelope shown in Figure $\ref{model}$ 
has been applied to analyze the observational results for 
c-C$_3$H$_2$ and CCH 
(Sakai et al. 2014a, 2014b). 
This model assumes that the gas motion follows the particle motion, 
which is governed by the gravity of the central mass with conservation of energy and angular momentum. 
A power-law distribution of the density is employed, 
and the intensity is assumed to be proportional to the column density along the line of sight. 
In spite of such a simple model, 
the basic features of the PV diagrams observed 
in L1527 for c-C$_3$H$_2$ and CCH are well reproduced. 
Therefore, it is worth applying the same model to the velocity structure observed for H$_2$CO and CCH in IRAS 15398--3359. 
In L1527, CCH resides only in infalling rotating the envelope. 
Hence, it is preferable to examine the toy model with the CCH line. 
However, the PV diagrams of CCH in IRAS 15398--3359 are complex due to the hyperfine structure, 
and their S/Ns are insufficient for critical comparison. 
In contrast, the emission of H$_2$CO ($5_{15}$--$4_{14}$) is bright enough in IRAS 15398--3359. 
Although H$_2$CO is present in the inner disk-like structure as well as the infalling rotating envelope in L1527, 
the latter contribution is significant (Sakai et al. 2014a). 
Hence, we compared the model results with the H$_2$CO ($5_{15}$--$4_{14}$) data of IRAS 15398--3359. 
Details of the calculation with the model is described in Appendix. 

\begin{figure}
	\includegraphics[bb = 0 0 100 650, scale = 0.8]{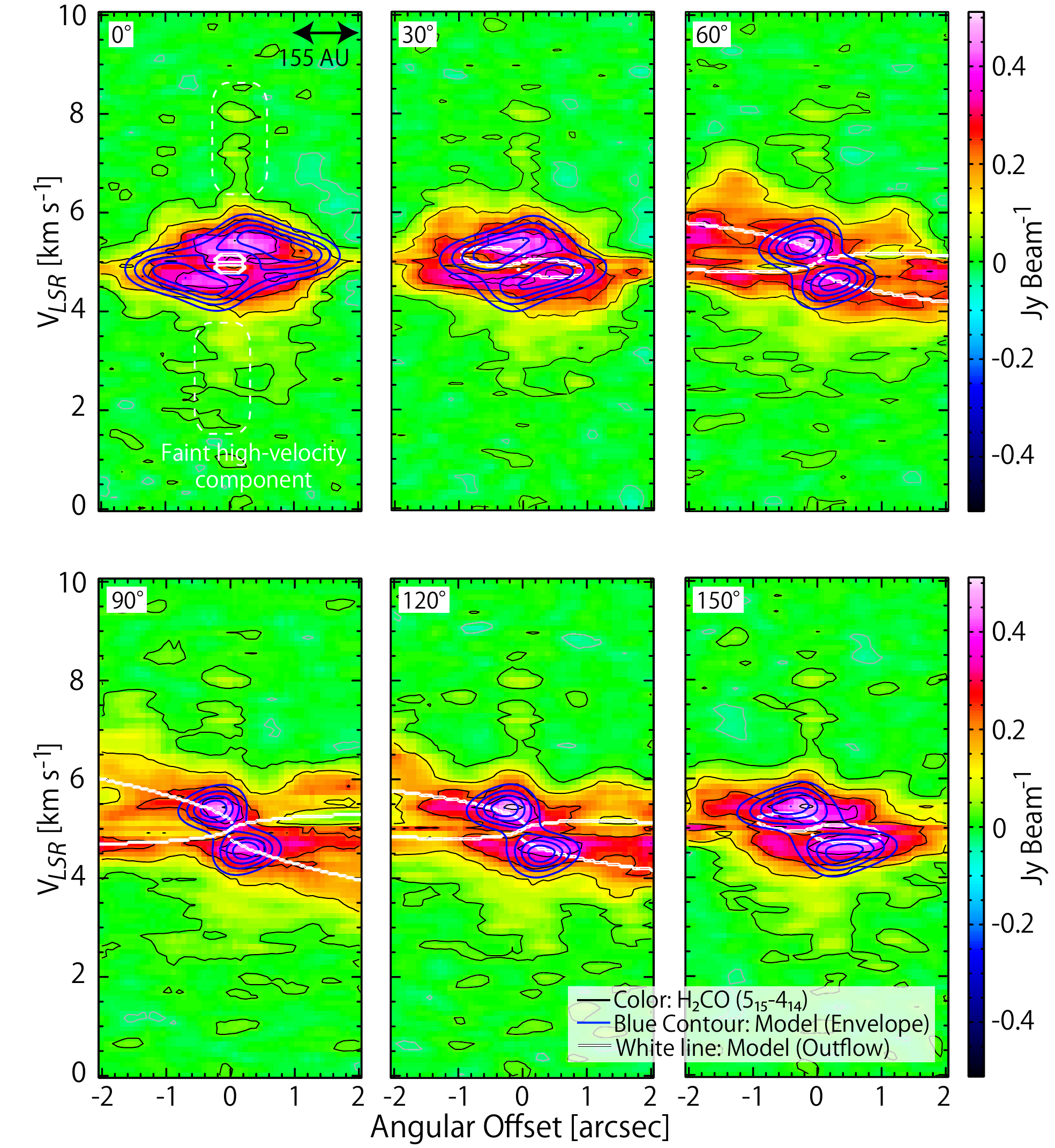}
	\caption{PV diagrams of H$_2$CO ($5_{15}$--$4_{14}$) along the six lines shown in Figure $\ref{H2CO_moment}$(f).  
			The blue contours show the model of the infalling rotating envelope. 
			The parameters for the model are 
			$I = 20\degr$, $M = 0.02\ M_\odot$, and $r_{\rm CB} = 30$ AU, 
			where $r_{\rm CB}$ is the radius of the centrifugal barrier. 
			The white lines show the model of the outflow cavity. 
			\label{H2CO_PV6}}
\end{figure}

Figure $\ref{H2CO_PVdiff}$ shows the PV diagrams of H$_2$CO ($5_{15}$--$4_{14}$) along the envelope axis. 
The blue contours represent the model results. 
Unlike in the L1527 case, we cannot determine the radius of the centrifugal barrier from the PV diagram 
because of absence of an obvious rotation signature. 
Nevertheless, 
we can roughly estimate its upper limit from the CCH distribution. 
If CCH is present only in the infalling rotating envelope as in the case of L1527 
and the radius of the centrifugal barrier is larger than the synthesized beam ($0\farcs5$; $78$ AU), 
a hole in the CCH distribution toward the protostar position should be resolved. 
Although an intensity dip toward the center is marginally seen in the CCH distribution (Figure $\ref{I-Profile}$), it is not well resolved in the present observation. 
Hence, we set the upper limit of the centrifugal radius to $80$ AU. 
We simulated the PV diagram for 
various sets of the protostellar mass and the radius of the centrifugal barrier under this constraint, 
as shown in Figure $\ref{H2CO_PVdiff}$. 
The models with a central mass of $0.02\ M_\odot$ and a radius of the centrifugal barrier of $0 - 30$ AU 
are chosen as the best simulations among the $20$ models shown in Figure $\ref{H2CO_PVdiff}$ by eye. 
The mass is consistent with the upper limit estimated in Section 3.3. 
Thus, the low protostellar mass can be confirmed with this simulation. 
We also simulated the PV diagrams along the different directions through the protostar position
shown in Figure $\ref{CCH_moment}$ (the six broken arrows labeled as ``$0\degr$", ``$30\degr$", ``$60\degr$", ``$90\degr$", ``$120\degr$", ``$150\degr$") 
for certain values of the central mass ($0.02\ M_\odot$) and the centrifugal barrier ($30$ AU) as an example
(Figure $\ref{H2CO_PV6}$).  
The trends of a velocity gradient around the protostar are reproduced by the model. 

\section{Discussion}
The low mass of the protostar IRAS 15398--3359 ($< 0.09\ M_\odot$) 
is essentially derived from the narrow line width in the inner envelope ($r < 80$ AU) 
which has an almost edge-on configuration. 
The above mass estimate may suffer from 
the uncertainty of the inclination angle of the envelope/disk. 
If the inclination angle is larger, 
the mass evaluated by the model becomes larger. 
However, the upper limit of the protostellar mass does not change significantly, 
even if an inclination angle of $30\degr$ is employed. 

By use of the dynamical timescale of the extended outflow ($1$--$2 \times 10^3$ yr for each lobe; U.A., Y\i ld\i z et al. submitted to A\&A) 
and the upper limit of the central mass ($< 0.09\ M_\odot$) obtained in this study, 
the average accretion rate is calculated to be less than  $9.0 \times 10^{-5}\ M_\odot$ yr$^{-1}$.  
The dynamical timescale of the outflow can be regarded as the lower limit, 
because the older part of the outflow may not be detected. 
Hence, the above accretion rate is regarded as the upper limit. 
Nevertheless, 
this estimate is 
roughly consistent with the typical accretion rate for low-mass protostars of ($10^{-5}$--$10^{-6}$) $M_\odot$ yr$^{-1}$ (e.g., Hartmann et al. 1997), 
and may be higher than 
that for another WCCC source L1527 ($10^{-6}\ M_\odot\ {\rm yr}^{-1}$; Ohashi et al. 1997). 
By use of the mass and the dynamical timescale of the outflow reported by van Kempen (2009), 
the mass outflow rate is calculated to be $6.3 \times 10^{-6}\ M_\odot\ {\rm yr}^{-1}$ 
and $2.2 \times 10^{-6}\ M_\odot\ {\rm yr}^{-1}$ for the red and blue lobes, respectively. 
These values correspond to 
$1.7 \times 10^{-5}\ M_\odot\ {\rm yr}^{-1}$ 
and $6.0 \times 10^{-6}\ M_\odot\ {\rm yr}^{-1}$, 
when an inclination angle of $20\degr$ is employed. 
Hence, the mass accretion rate and the mass outflow rate are roughly comparable to each other. 

On the other hand, 
the mass accretion rate can be estimated by use of the equation (Palla \& Stahler 1991): 
\begin{equation}
	\dot{M} = \frac{L R_{\rm star}}{G M}, 
\end{equation}
where the $L$ is the luminosity and the $R_{\rm star}$ the radius of the protostar. 
By using the obtained upper limit to $M$ ($0.09\ M_\odot$), 
the mass accretion rate $\dot{M}$ is estimated to be larger than 
$1.6 \times 10^{-6}\ M_\odot\ {\rm yr}^{-1}$ 
with $L = 1.8\ L_\odot$ (J\o rgensen et al. 2013) and 
$R_{\rm star} = 2.5\ R_\odot$ (e.g., Palla 1999; Baraffe \& Chabrier 2010). 
This lower limit is consistent with the abovementioned upper limit 
($9.0 \times 10^{-5}\ M_\odot\ {\rm yr}^{-1}$). 
In spite of the episodic accretion suggested by J\o rgensen et al. (2013), 
the average mass accretion rate is not very different from the canonical value 
($10^{-5}$--$10^{-6}\ M_\odot\ {\rm yr}^{-1}$).  

On the other hand, 
the upper limit of the protostar/envelope mass ratio is evaluated to be $0.18$ 
from the upper limit of the protostellar mass of $0.09\ M_\odot$ and an envelope mass of $0.5\ M_\odot$ (Kristensen et al. 2012). 
The ratio is $0.04$, if the 
mass of $0.02\ M_\odot$ is employed. 
Hence, the ratio seems smaller than that in L1527 ($0.2$; Tobin et al. 2012). 
This implies that the protostar is in the infant stage and is still growing. 
Since the specific angular momentum brought into the inner region is smaller in the earlier stage of the protostellar evolution, 
it seems likely that the rotation signature is not so evident as in the L1527 case. 

The H$_2$CO distribution has a centrally concentrated component with a single-peaked distribution, 
whereas the CCH distribution is more flattened (Figure $\ref{I-Profile}$). 
Therefore, the H$_2$CO line traces the inner region in comparison with the CCH line. 
This trend is also seen in the observations of L1527 (Sakai et al. 2014a). 
It should be noted that 
Figure $\ref{H2CO_PV6}$ shows a faint high-velocity component of H$_2$CO 
($v_{\rm lsr} < 4$ km s$^{-1}$ or $v_{\rm lsr} > 6$ km s$^{-1}$) toward the protostar position. 
If this component really represents the contribution of a rotationally supported disk, 
it suggests that a disk structure is formed already in the early stages of the low-mass star formation. 
If the upper limit of the central mass of $0.09\ M_\odot$ is employed, 
H$_2$CO may be present even at $\sim 10$ AU, 
according to its maximum velocity shift ($\sim 3$ km s$^{-1}$). 
We here stress that extensive studies of the disk structure in Class 0 sources 
have now become possible with ALMA 
and that a chemical approach will be of help to such studies. 
Our result reveals that the disk/envelope structure can be observed with various molecular lines. 
Hence, the chemical evolution can be investigated even in the disk-forming stage. 

\acknowledgments
This paper makes use of the following ALMA data set ADS/JAO.ALMA$\#$2011.0.00777.S. 
ALMA is a partnership of the ESO (representing its member states), 
the NSF (USA) and NINS (Japan), together with the NRC (Canada) and the NSC and ASIAA (Taiwan), 
in cooperation with the Republic of Chile. 
The Joint ALMA Observatory is operated by the ESO, the AUI/NRAO and the NAOJ. 
The authors are grateful to the ALMA staff for their excellent support. 
N.S. and S.Y. acknowledge financial support from Grant-in-Aid from the Ministry of 
Education, Culture, Sports, Science, and Technologies of Japan (21224002, 25400223, and 25108005). 
Y.O. thanks the Advanced Leading Graduate Course for Photon Science (ALPS) for financial support.  

\clearpage

\appendix
\section{Configuration of the Envelope Model}
\begin{figure}
	\includegraphics[bb = -50 0 100 480, scale = 1.0]{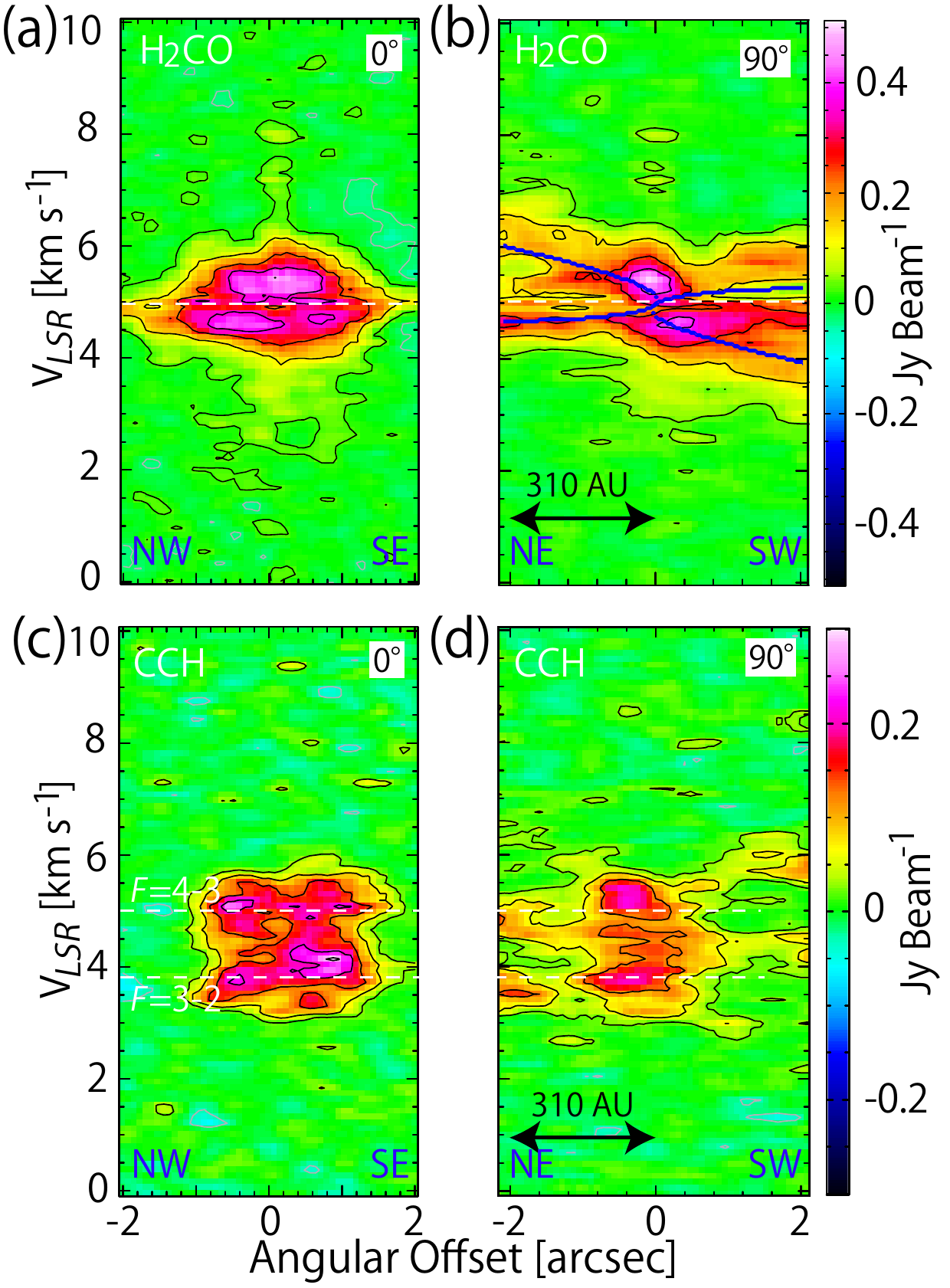}
	\caption{PV diagrams of H$_2$CO ($5_{15}$--$4_{14}$) ((a) and (b)) and CCH ($F_2$) ((c) and (d)). 
			The diagrams of panels (a) and (c) are constructed  
			along the line perpendicular to the outflow axis, 
			while those of panels (b) and (d) along the outflow axis, 
			as shown in Figure $\ref{H2CO_moment}$(f)  
			(the black broken arrows labeled as ``$0\degr$" and ``$90\degr$" respectively). 
			Diagrams (b) and (d) are the zooms of the central parts 
			of Figures $\ref{PV_forI}$(a) and $\ref{PVoutflow}$, respectively. 
			The blue lines in the diagram of panel (b) show the model of the outflow. 
			\label{PV2x2}}
\end{figure}

\begin{figure}
	\includegraphics[bb = -20 0 100 150, scale = 0.9]{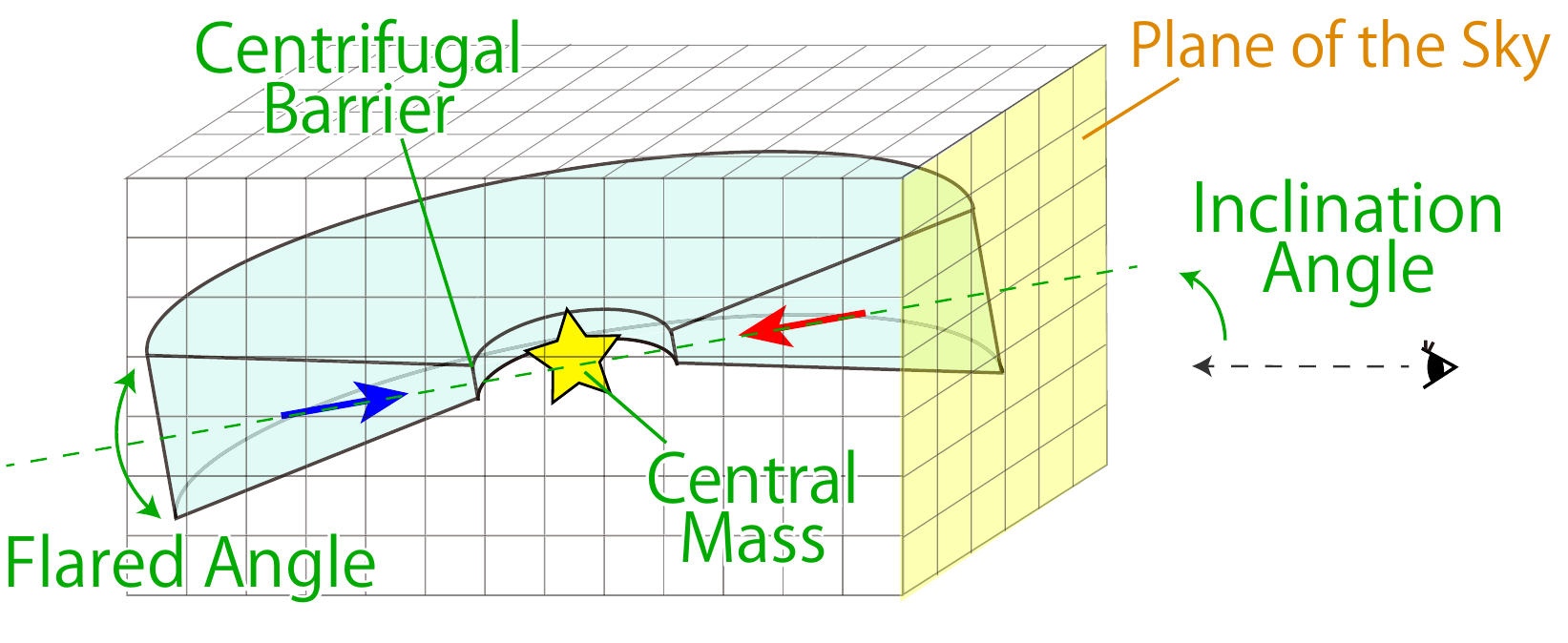}
	\caption{Model of the infalling rotating envelope. 
			The distribution and the velocity field of the gas are sectioned into meshes. 
			\label{mesh}}
\end{figure}

In our envelope model, 
the three-dimensional space is sectioned into meshes centered around the protostar, 
as shown in Figure $\ref{mesh}$, 
and each mesh has data of the abundance of the gas and the velocity field. 
The intensities of the emission from the meshes are calculated under the assumptions as follows. 
\vspace{-15pt}
\begin{itemize}
\item The motion of the gas is approximated by the particle motion, where the gas pressure is neglected. 
	The particle motion is governed only by the gravity of the central mass 
	with conservation of its energy and angular momentum. 
\item The envelope has a flared shape with a flared angle of $20\degr$. 
	Its outer radius is fixed to be $240$ AU, which is the radius shown in Figure $\ref{PV2x2}$(a). 
\item	Although the envelope is not spherical,
	the density of the gas in the envelope is assumed to be proportional to $r^{-1.5}$ 
	outside the centrifugal barrier for simplicity 
	(e.g., Shu 1977; Ohashi et al. 1997; Harvey et al. 2003). 
	This corresponds to the density profile of an infalling cloud. 
	However, this choice is rather arbitrary in this study,  
	because our main interest in this model is on the velocity field of the gas around the protostar. 
	Furthermore, we assume that the gas is absent inward of the centrifugal barrier. 
	Namely, no rotationally supported disk structure is considered. 
	This can be justified by a lack of the clear Keplerian rotation feature in the PV diagram.  
\item 
	The molecular emission is simply assumed to be optically thin. 
	Therefore, the intensity is just proportional to the number of molecules along the line of sight. 
	No radiative transfer effects are considered. 
\item The spectral line has a Gaussian profile with a line width of $0.5$ km s$^{-1}$, 
	and the emission is observed 
	with a Gaussian beam with a FWHM of $0\farcs6 \times 0\farcs6$. 
\item The number of the meshes is fixed to be ${\displaystyle \left(2^7 + 1\right)^3}$ 
	and each mesh is a cube with a size of $0\farcs04$ ($= 6.2$ AU). 
	The velocities of the gas are discretized with a step of $0.04$ km s$^{-1}$. 
\end{itemize}
\vspace*{-15pt}
The model parameters are the inclination angle $i$, 
the central mass $M$, and the radius of the centrifugal barrier $r_{\rm CB}$. 
The radius of the centrifugal barrier is represented as 
\begin{align*}
	r_{\rm CB} &= \frac{1}{2GM} \left( \frac{L}{m} \right)^2, 
\end{align*}
where $G$ is the gravitational constant. 
It is the radius at which all the kinetic energy is converted to the rotational energy, 
and is determined by the relative strength of the gravitational force and the centrifugal force. 
Hence, specification of $r_{\rm CB}$ and $M$ means specification of the specific angular momentum ${\displaystyle \frac{L}{m}}$. 
The infalling and rotating velocities of particles at the distance $r$ to the protostar are represented as follows: 
\begin{align*}
	v_{\rm rotation} &= \left( \frac{L}{m} \right) \frac{1}{r}, \\ 
	v_{\rm infall} &= \sqrt{\frac{2GM}{r} - v_{\rm rotation}^2}. 
\end{align*}


\clearpage

\clearpage

\clearpage

\clearpage

\clearpage

\clearpage

\clearpage
\begin{figure}
	\includegraphics[bb = 0 0 100 600, scale = 0.6]{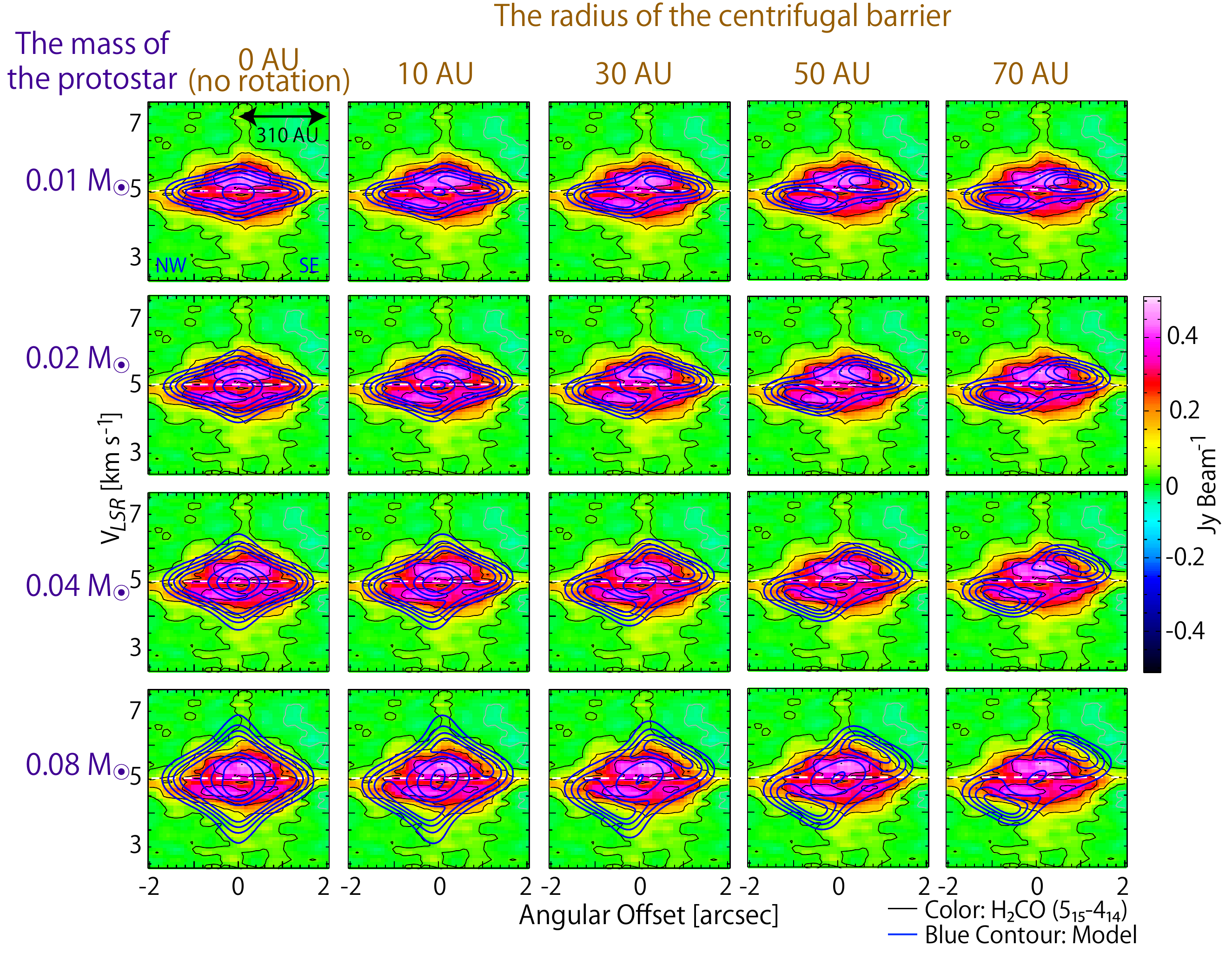}
	\caption{PV diagrams of H$_2$CO ($5_{15}$--$4_{14}$) 
			along the line perpendicular to the outflow axis (``$0\degr$") shown in Figure $\ref{H2CO_moment}$(f). 
			The blue contours represent 16 model simulations of the infalling rotating envelope 
			with an inclination angle of $20\degr$. 
			The model parameters are the mass of the protostar and the radius of the centrifugal barrier (see Appendix). 
			\label{H2CO_PVdiff}}
\end{figure}

\clearpage

\clearpage

\clearpage
\begin{table}
	\caption{Observed Line Parameters\tablenotemark{a} \label{line}}
	\begin{center}
		\begin{tabular}{lccc}
			\hline \hline 
			Transition & Frequency & $E_{\rm u} k^{-1}$ & S$\mu^2\ \tablenotemark{b}$ \\ 
			 & (GHz) & (K) & (D$^2$) \\ 
			 \hline
			 \multicolumn{4}{c}{H$_2$CO} \\
			 \hline 
			 $5_{05}$--$4_{04}$ & 362.7360480 & 42 & 27.168 \\ 
			 $5_{15}$--$4_{14}$ & 351.7686450 & 49 & 26.096 \\ 
			 $5_{24}$--$4_{23}$ & 363.9458940 & 75 & 22.834 \\ 
			 \hline
			 \multicolumn{4}{c}{CCH} \\
			 \hline
			 $N$=4--3, $J$=7/2--5/2, $F$=3--2 & 349.4006712 & 34 & 1.6942 \\ 
			 $N$=4--3, $J$=7/2--5/2, $F$=4--3 & 349.3992756 & 34 & 2.2712 \\ 
			 \hline
		\end{tabular}
	\end{center}
	\tablenotetext{a}{Taken from CDMS (M\"{u}ller et al. 2005).}
	\tablenotetext{b}{Nuclear spin degeneracy is not included.}
\end{table}
\end{document}